\documentclass[conference,a4paper]{APSIPA2025}
\usepackage{amsmath}
\usepackage{graphicx}
\usepackage{multirow}
\usepackage{subfigure}
\usepackage{threeparttable}
\usepackage[
  style=ieee,
  backend=biber,
  doi=false,
  isbn=false,
  url=false,
  eprint=false
]{biblatex}
\usepackage{parskip}
\usepackage{geometry}
\usepackage{fancyhdr}

\fancypagestyle{firststyle}{
  \fancyhf{}
  \fancyhead[C]{2025 Asia Pacific Signal and Information Processing Association Annual Summit and Conference (APSIPA ASC)}
}

\begin{document}

\title{Hierarchical Sparse Sound Field Reconstruction with Spherical and Linear Microphone Arrays}

\author{
\authorblockN{
Shunxi Xu and
Craig T. Jin
}

\authorblockA{
The University of Sydney, Sydney, Australia \\
E-mail: {shunxi.xu, craig.jin}@sydney.edu.au}}

\maketitle
\thispagestyle{firststyle}
\pagestyle{fancy}

\begin{abstract}
Spherical microphone arrays (SMAs) are widely used for sound field analysis, and sparse recovery (SR) techniques can significantly enhance their spatial resolution by modeling the sound field as a sparse superposition of dominant plane waves. However, the spatial resolution of SMAs is fundamentally limited by their spherical harmonic order, and their performance often degrades in reverberant environments. This paper proposes a two-stage SR framework with residue refinement that integrates observations from a central SMA and four surrounding linear microphone arrays (LMAs). The core idea is to exploit complementary spatial characteristics by treating the SMA as a primary estimator and the LMAs as a spatially complementary refiner. Simulation results demonstrate that the proposed SMA–LMA method significantly enhances spatial energy map reconstruction varying reverberation conditions, compared to both SMA-only and direct one-step joint processing. These results demonstrate the effectiveness of the proposed framework in enhancing spatial fidelity and robustness in complex acoustic environments.
\end{abstract}

\section{Introduction}
\label{sec:intro}

Spherical microphone arrays (SMAs) have become essential tools for capturing spatial audio and analyzing acoustic sound fields. A key advantage of SMAs is that they provide a panoramic point of view on the sound field, enabling a wide variety of spatial analysis techniques. For instance, simple beamforming methods can provide an energy map of the sound field \cite{park2005sound}. There are also more elaborate techniques such as multiple signal classification (MUSIC)  and the estimating signal parameters via rotational invariance technique (ESPRIT), which have recently been applied to SMAs, which are referred to as EB-MUSIC \cite{cohen2009speech} and EB-ESPRIT \cite{5946342}.

Sparse recovery (SR) techniques have recently gained significant attention as effective methods to enhance the spatial resolution of SMA recordings \cite{jin2020perspectives}. By representing the sound field as a superposition of a limited number of dominant plane-wave components, SR-based methods provide spatial resolutions surpassing traditional beamforming techniques, thereby enabling finer spatial detail and improved localization accuracy. To further improve robustness and convergence in practical acoustic environments, several enhancements have been proposed, including independent component analysis (ICA) with compressive sensing (CS) fusion \cite{noohi2013direction}, subspace pre-processing \cite{epain2013super}, and spatial priming \cite{noohi2015super}.

Despite these advances, SR-based SMA methods remain constrained by the finite spherical harmonic (SH) order, which limits spatial resolution and introduces aliasing at higher frequencies \cite{jin2013design}. Moreover, SR methods hold the assumption that the sound field comprises only a few dominant plane-wave components, which is often violated in real-world conditions with significant reverberation, diffuse noise, or densely populated acoustic environments \cite{wu2012dereverberation}. Even advanced pre-processing and spatial priming techniques are constrained by the intrinsic resolution limits of SMA data in certain frequency ranges and directions.

To complement SMA’s limitations, linear microphone arrays (LMAs) can provide additional spatial detail due to their high directional sensitivity along the array axis. Despite their simple and compact geometry, LMAs are particularly effective in resolving closely spaced sources in moderately reverberant environments \cite{hu2014near}. To further mitigate the inherent spatial ambiguities of linear arrays, configurations with multiple LMAs arranged along the x- and y-axes have been proposed to enhance horizontal localization while preserving hardware simplicity \cite{chen2005robust}. Previous research has extended this idea to multiple SMAs by simply concatenating their observations into a one-step SR framework \cite{tang2022wave}, but no work has addressed SMA–LMA combinations. However, such direct concatenation proves ineffective in practice, as the performance of LMAs degrades severely in reverberant environments, ultimately compromising the outcome of the joint processing.


In this paper, we propose a two-stage SR framework that treats a central SMA as the primary estimator and the four surrounding LMAs as a spatially complementary refiner. This structure preserves the high-quality spatial information extracted by the SMA while isolating it from the limitations of the LMAs, which could otherwise degrade performance due to elevation ambiguity and reverberation sensitivity. In the first stage, SR is performed in the SH domain using SMA observations to obtain an initial estimate of the sound field. This estimate is projected into the LMA domain and subtracted from the actual LMA measurements to isolate a residue signal that captures spatial components unresolved by the SMA. A second SR stage is then applied to the residue using the LMA model. The final spatial map is constructed by fusing the SMA estimate and the LMA refinement, resulting in substantially improved spatial map reconstruction. 

The remainder of this paper is organized as follows. Section 2 reviews sparse plane-wave decomposition and SR techniques for SMAs and LMAs. Section 3 details the proposed joint SMA-LMA residue refinement method. Section 4 presents simulation results in reverberant environments, and Section 5 concludes the paper and outlines future work.

\section{Background}
\label{sec:background}

\subsection{Sparse Plane-Wave Decomposition}
\label{subsec: Sparse Plane Wave Decomposition}
In classic plane-wave decomposition, the sound field is modeled as a superposition of $N$ plane waves arriving from many directions in space. Assume the sound field is observed by an array providing $K$ signals, with $K \ll N$. Using a short-time Fourier transform (STFT) domain representation, the relationship between the observed microphone signals $\mathbf{B}(t,f)$ and the plane-wave source signals $\mathbf{X}(t,f)$ can be expressed as:
\begin{equation}
    \mathbf{B}(t,f) = \mathbf{D}(f)\,\mathbf{X}(t,f),
    \label{eq:plane_wave_decomposition}
\end{equation}
where $t$ denotes the time frame index, $f$ is the frequency bin index, and the dictionary matrix $\mathbf{D}(f)$ characterizes how each plane wave contributes to the array observations. The dictionary matrix is explicitly defined as:
\begin{equation}
\begin{aligned}
    &\mathbf{D}(f) = [\mathbf{d}(\mathbf{\Omega}_1,f), \mathbf{d}(\mathbf{\Omega}_2,f),\dots,\mathbf{d}(\mathbf{\Omega}_N,f)],\\
    &\mathbf{d}(\mathbf{\Omega}_n,f) = [d_1(\mathbf{\Omega}_n,f), d_2(\mathbf{\Omega}_n,f),\dots,d_K(\mathbf{\Omega}_n,f)]^T,
\end{aligned}   
\end{equation}
where column vector $\mathbf{d}(\mathbf{\Omega}_n,f)$ is termed a manifold vector, describing the complex-valued transfer function from a plane wave originating from direction $\mathbf{\Omega}_n$ at frequency $f$ to each array microphone.

Recovering the sound field $\mathbf{X}(t,f)$ from the observations $\mathbf{B}(t,f)$ involves solving the inverse problem posed by \eqref{eq:plane_wave_decomposition}, which is typically underdetermined, making direct inversion methods ill-posed and unstable \cite{candes2008introduction}. The sparse plane-wave decomposition method seeks the solution with the fewest plane waves that still explains the observations \cite{jin2017sound}. The sparse solution with the fewest plane waves can be obtained by solving the following constrained optimization problem:
\begin{equation}
    \hat{\mathbf{X}}(t,f) = \underset{\mathbf{X}(t,f)}{\text{argmin}}\;||\mathbf{X}(t,f)||_{2,p}\,\,\,\,\text{s.t.}\,\,\,\,\mathbf{B}(t,f) = \mathbf{D}(f)\,\mathbf{X}(t,f),
    \label{eq:sparse_optimization_problem}
\end{equation}
where the mixed $\ell_{2,p}$-norm $||\cdot||_{2,p}$ promotes sparsity in the solution. It is explicitly defined as:
\begin{equation}
    ||\mathbf{X}||_{2,p} = \left(\sum_{i=1}^N \left(\sqrt{\sum_{t=1}^T |x_i(t,f)|^2}\right)^p\right)^{\frac{1}{p}},\quad 0 < p < 1,
\end{equation}
where the summation over $t$ aggregates across time frames, ensuring that sparsity is enforced jointly over the temporal dimension for each plane wave direction.

In this work, the sparse optimization problem in \eqref{eq:sparse_optimization_problem} is solved using the Iteratively Reweighted Least Squares (IRLS) algorithm, as detailed in \cite{daubechies2010iteratively}. The estimated $\hat{\mathbf{X}}(t,f) $ is also referred to as the SR coefficients that describe how the plane-waves explain the observed sound field.

\subsection{SMA Processing in the SH Domain}
\label{subsec:SMA Processing in the SH Domain}
The sound pressure, $p(k, r, \theta, \phi)$, measured on the surface of a spherical array (open or rigid) with spherical coordinates $(r,\theta,\phi)$ at the frequency $f$ can be modeled as a summation of $L$ spherical modes \cite{rafaely2015fundamentals}:  
\begin{equation}
p(k, r, \theta, \phi) =
\sum_{n=0}^{L} \sum_{m=-n}^{n}
a_{nm}(k)\, b_n(kr)
\,Y_n^m(\theta, \phi),
\end{equation}
where $k = 2\pi f / c$ is the wave number with $c$ denoting the speed of sound, and $Y_n^m(\theta, \phi)$ are the spherical harmonic functions of order $n$ and degree $m$. The coefficients $a_{nm}(k)$ represent the spherical harmonic expansion of the sound field, and $b_n(kr)$ is the so-called mode strength, given by:
\begin{equation}
b_n(kr) = 4\pi i^n
\begin{cases}
 j_n(kr), & \text{open sphere} \\
j_n(kr) - \frac{j_n'(k r_a)}{h_n^{(2)\prime}(k r_a)} h_n^{(2)}(kr), & \text{rigid sphere}
\end{cases}
\end{equation}
where $j_n$ is the spherical Bessel function of order $n$. $h_n^{(2)}$ is the spherical Hankel function of the second kind and of order $n$. $j_n'$ and $h_n'$ denote their first derivatives. Then the pressure in the SH domain can be written as
\begin{equation}
    p_{nm}(k,r) = a_{nm}(k)\, b_n(kr).
\end{equation}
In the sparse plane-wave decomposition framework \eqref{eq:plane_wave_decomposition}, the observations in the SH domain with time index $n$ and frequency index $f$ can be expressed as
\begin{equation}
    \mathbf{B}_{\text{HOA}}(t,f) = [p_{00}(t,f),p_{1-1}(t,f),\dots,p_{LL}(t,f)]^T,
\end{equation}
and the dictionary matrix $\mathbf{D}_{\text{HOA}}$ representing $N$ possible plane-wave directions in the SH domain is defined as
\begin{equation}
\begin{aligned}
    &\mathbf{D}_{\text{HOA}} = [\mathbf{d}(\theta_1, \phi_2),\mathbf{d}(\theta_2, \phi_2),\dots,\mathbf{d}(\theta_n, \phi_n)],\\
    &\mathbf{d}(\theta_i, \phi_i) = [Y_0^0(\theta_i, \phi_i),Y_1^{-1}(\theta_i, \phi_i),\dots,Y_L^L(\theta_i, \phi_i)]^T,
\end{aligned}
\end{equation}
where $\mathbf{d}(\theta_i, \phi_i)$ is a vector giving the SH components of a plane-wave for direction $(\theta_i, \phi_i)$. Note that the dictionary is frequency independent in the SH domain, as the SH functions are frequency independent. Given the dictionary and the observations, the sparse plane-wave decomposition in the SH domain is obtained by solving the optimization problem in \eqref{eq:sparse_optimization_problem}, finding the direction-dependent plane-wave coefficients $\hat{\mathbf{X}}_{\text{HOA}}(t,f)$ that best explain the observed sound field.

\subsection{LMA Processing in the Signal Domain}
\label{subsec:LMA_Processing}

LMAs are typically processed directly in the signal (time-frequency) domain. The sound pressure signal received at the $q$-th microphone at time $t$ and frequency $f$ can be modeled as a superposition of $N$ plane-wave sources:
\begin{equation}
p_q(t,f) = \sum_{n=1}^{N} s_n(t,f)\, e^{-j\frac{2\pi f}{c} \langle \mathbf{r}_q, \mathbf{u}_n \rangle} + w_q(t,f),
\end{equation}
where $s_n(t,f)$ denotes the signal of the $n$-th source, $\mathbf{r}_q$ is the position vector of the $q$-th microphone, $\mathbf{u}_n$ is the unit direction vector of the $n$-th source, and $w_q(t,f)$ represents additive noise.

Following the sparse decomposition framework, the observation matrix from $Q$ linearly placed microphones  $\mathbf{B}(t,f)$ is constructed as
\begin{equation}
    \mathbf{B}_{\text{LMA}}(t,f) = [p_1(t,f),p_2(t,f),\dots,p_Q(t,f)]^T,
\end{equation}
and the dictionary matrix $\mathbf{D}_{LMA}(f)$ composed of $N$ steering vectors corresponding to $N$ possible directions is defined as
\begin{equation}
\begin{aligned}
    &\mathbf{D}_{\text{LMA}}(f) = [\mathbf{d}(\mathbf{u}_1,f),\mathbf{d}(\mathbf{u}_2,f),\dots,\mathbf{d}(\mathbf{u}_N,f)],\\
    &\mathbf{d}(\mathbf{u}_n,f) = [e^{-j\frac{2\pi f}{c} \langle \mathbf{r}_1, \mathbf{u}_n \rangle},e^{-j\frac{2\pi f}{c} \langle \mathbf{r}_2, \mathbf{u}_n \rangle},\dots,e^{-j\frac{2\pi f}{c} \langle \mathbf{r}_Q, \mathbf{u}_n \rangle}]^T,
\end{aligned}
\end{equation}

where $\mathbf{d}(\mathbf{u}_n,f) \in \mathcal{C}^{Q}$ is the steering vector corresponding to direction $\mathbf{u}_n$. Given the dictionary and the observations, the SR coefficients $\hat{\mathbf{X}}_{\text{LMA}}(t,f)$ in the signal domain is obtained by solving the optimization problem in \eqref{eq:sparse_optimization_problem}. 

\subsection{Joint Processing for SMA and LMA}
\label{subsec:joint processing}
A basic approach to jointly exploit observations from multiple arrays is to concatenate their respective dictionary matrices and perform a single-stage SR. Specifically, given the SMA observation $\mathbf{B}_{\text{HOA}}(t,f)$ in the SH domain and the LMA observation $\mathbf{B}_{\text{LMA}}(t,f)$ in the signal domain, the observation matrix is defined as:
\begin{equation}
\mathbf{B}_{\text{concat}}(t,f) = 
\begin{bmatrix}
\mathbf{B}_{\text{HOA}}(t,f) \\
\mathbf{B}_{\text{LMA}}(t,f)
\end{bmatrix},
\end{equation}
and the corresponding concatenated dictionary as:
\begin{equation}
\mathbf{D}_{\text{concat}}(f) = 
\begin{bmatrix}
\mathbf{D}_{\text{HOA}}(f) \\
\mathbf{D}_{\text{LMA}}(f)
\end{bmatrix}.
\end{equation}

Solving the optimization problem in \eqref{eq:sparse_optimization_problem} yields the estimated SR coefficients $\hat{\mathbf{X}}_{\text{concat}}(t,f)$, which represent plane-wave components inferred from concatenated SMA and LMA observations, assuming implicit equal weighting. However, this direct fusion approach is highly sensitive to reverberation. Reflections captured by the LMA can introduce spurious components that distort the sparse recovery process and compromise reconstruction accuracy. This motivates the more resilient strategy introduced in the next section.

\section{Proposed Method for Joint SMA and LMA Processing}
\label{sec:proposed_method}

This section presents the proposed two-stage SR framework with residue refinement (RR) for joint processing of SMA and LMA observations to improve SR-based sound field analysis in reverberant conditions. The idea is to leverage the high spatial resolution of the SMA for the initial reconstruction and then refine the estimate by resolving residual components using LMA data. Fig.\,\ref{fig:flowchart} illustrates the procedure of the proposed method, with detailed steps described below:
\begin{figure}[t]
  \centering
  \centerline{\includegraphics[width=0.9\columnwidth]{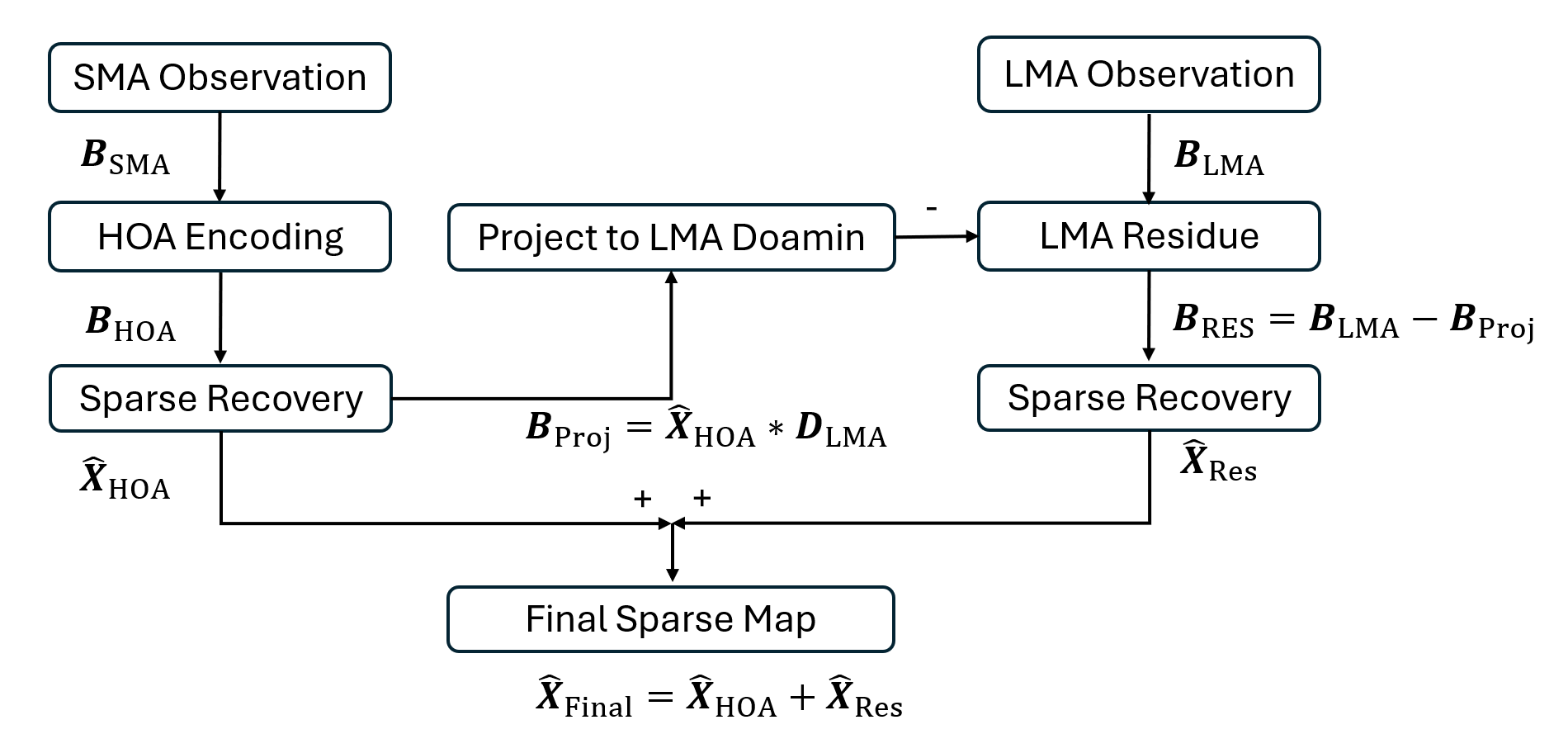}}
  \caption{Block diagram of the proposed joint SMA-LMA residue refinement method.}
  \label{fig:flowchart}
\end{figure}

\begin{enumerate}
    \item \textbf{Initial Sparse Recovery on SMA observations}: The SMA signals are first transformed into the SH domain using HOA encoding. Sparse recovery is then applied to obtain the SR coefficients:
    \begin{equation}
    \begin{aligned}
        \hat{\mathbf{X}}_{\text{HOA}}(t,f) = \underset{\mathbf{X}_{\text{HOA}}(t,f)}{\text{argmin}}\;||\mathbf{X}_{\text{HOA}}(t,f)||_{2,p}\quad \\ \text{s.t.}\quad\mathbf{B}_{\text{HOA}}(t,f) = \mathbf{D}_{\text{HOA}}(f)\,\mathbf{X}_{\text{HOA}}(t,f). \quad \quad 
    \end{aligned}
    \end{equation}
    
    \item \textbf{Projection and Residue Calculation in LMA domain}: The SMA-based estimate is projected to the LMA domain by using the dictionary of LMA under the plane-wave assumption:
    \begin{equation}
        \mathbf{B}_{\text{Proj}}(t,f) = \mathbf{D}_{\text{LMA}}(f)\,\hat{\mathbf{X}}_{\text{HOA}}(t,f).
    \end{equation}
    The difference between the actual LMA observations and this projection defines the residue, representing the part of the sound field that was not captured by the SMA:
    \begin{equation}
        \mathbf{B}_{\text{Res}}(t,f) = \mathbf{B}_{\text{LMA}}(t,f) - \mathbf{B}_{\text{Proj}}(t,f).
    \end{equation}

    \item \textbf{Sparse Recovery on Residue}: A second sparse recovery stage is applied to the residue using the LMA model to extract the missing spatial components that were not recovered in the initial SMA stage:
    \begin{equation}
    \begin{aligned}
        \hat{\mathbf{X}}_{\text{Res}}(t,f) = \underset{\mathbf{X}_{\text{Res}}(t,f)}{\text{argmin}}\;||\mathbf{X}_{\text{Res}}(t,f)||_{2,p}\quad \\ \text{s.t.}\quad\mathbf{B}_{\text{Res}}(t,f) = \mathbf{D}_{\text{LMA}}(f)\,\mathbf{X}_{\text{Res}}(t,f). \quad \quad 
    \end{aligned}
    \end{equation}

    \item \textbf{Fusion of SMA and LMA Estimates}: The final sparse estimate is obtained by summing the SMA-based and LMA-refined coefficients:
    \begin{equation}
        \hat{\mathbf{X}}_{\text{Final}}(t,f) = \hat{\mathbf{X}}_{\text{HOA}}(t,f) + \hat{\mathbf{X}}_{\text{Res}}(t,f).
    \end{equation}
\end{enumerate}

This fusion allows the LMA to contribute unresolved spatial components, particularly along its sensitive axis. Depending on the LMA geometry, some ambiguity may arise in off-axis and elevation directions due to limited aperture and spatial symmetries. However, these effects are constrained to the residue, while the SMA provides a stable and accurate 3D estimate in the first stage. This hierarchical design ensures that the high-quality global structure recovered by the SMA is preserved and protected from degradation during refinement.

Once the plane-wave signal decomposition has been obtained, an acoustic energy map can be constructed to visualize the spatial distribution of incoming acoustic energy. For each direction 
$\mathbf{\Omega}_n$ represented in the plane-wave dictionary, the corresponding energy 
$e(\mathbf{\Omega}_n)$ is computed by summing the squared magnitude of the estimated SR coefficients across all time frames and frequency bins:
\begin{equation}
    e(\mathbf{\Omega}_n) =\sum_{t=1}^{T}\sum_{f=1}^{F}|\hat{\mathbf{X}}_{\text{Final},n}(t,f)|^2,
\end{equation}
where $T$ and $F$ denote the total number of time frames and frequency bins, respectively.
Clearly, the spatial resolution of the resulting energy map is determined by the angular resolution of the plane-wave dictionary used in the sparse recovery process.

\begin{figure}[t]
  \centering
  \centerline{\includegraphics[width=0.75\columnwidth]{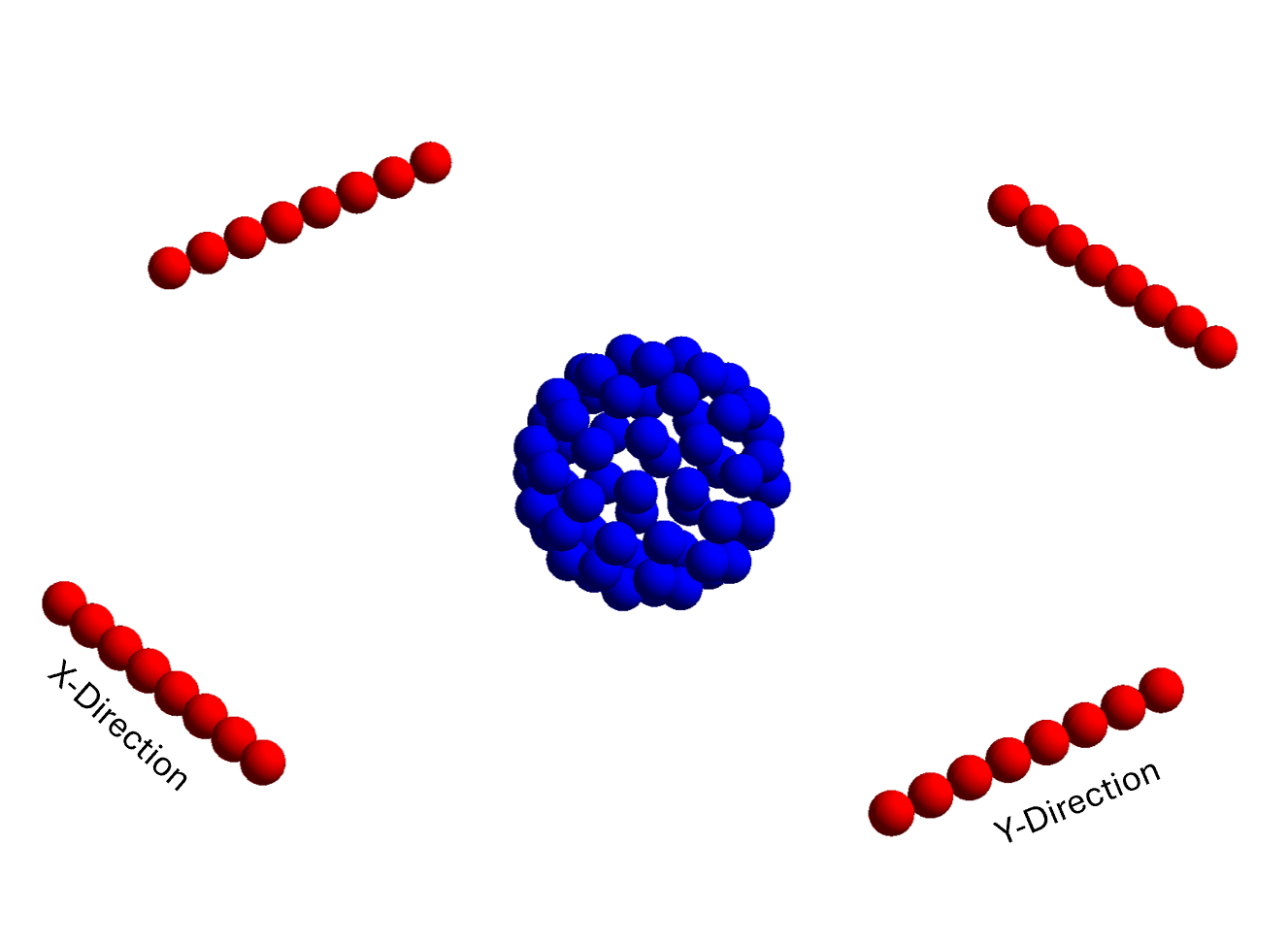}}
  \caption{Microphone array configuration used in the simulation. Blue dots represent the 64-element open SMA, and red dots denote the 4 surrounding 8-element LMAs.}
  \label{fig:array setup}
\end{figure}

\section{Numerical Simulations}
\label{sec:numerical simulations}

\subsection{Simulation Setup}
\label{subsec:simulation setup}

To evaluate the proposed residue refinement framework, we simulate microphone array recordings in reverberant environments using the MCRoomSim room impulse response (RIR) generator \cite{wabnitz2010room}. The simulated room measures $10\,\text{m} \times 8\,\text{m} \times 3\,\text{m}$ with a reverberation time (RT60) of $0.3\,\text{s}$. Two array configurations are tested: (i) a single open SMA with 64 omnidirectional microphones uniformly distributed on a  $10\,\text{cm}$ radius sphere, placed at the center of the room, and (ii) a hybrid setup combining the same SMA with four surrounding LMAs. Each LMA contains 8 microphones with a spacing of $4\,\text{cm}$, and they are symmetrically placed $0.5\,\text{m}$ from the SMA center\,—\,two aligned along the x-axis and two along the y-axis, forming a square-like layout as illustrated in Fig.\,\ref{fig:array setup}.

Dry speech signals of $4\,\text{s}$ duration are used as clean sources, simulated as plane waves arriving from random directions at three distances from the SMA center: $1.5\,\text{m}$, $2.5\,\text{m}$, and $3.5\,\text{m}$. The number of active sources ranges from 2 to 10, with 100 independent trials conducted for each case using randomly sampled source directions. Microphone signals are generated by convolving the clean speech with the corresponding RIRs and adding spatially uncorrelated white Gaussian noise to achieve an SNR of $30\,\text{dB}$.

The SMA signals are encoded into the HOA domain up to order 4. Sparse recovery is carried out using the IRLS algorithm. The regularization parameter is estimated dynamically using the diffuseness of the HOA signals \cite{epain2016spherical}. The IRLS optimization starts with $\ell_1$-norm minimization for the first 10 iterations, and then switches to $\ell_p$-norm minimization with $p = 0.7$ for improved sparsity as suggested in \cite{daubechies2010iteratively}. The SR dictionary $\mathbf{D}$ is constructed from 642 uniformly sampled directions obtained by recursively subdividing an icosahedron.

\begin{figure}[t]
  \centering
  \centerline{\includegraphics[width=1\columnwidth]{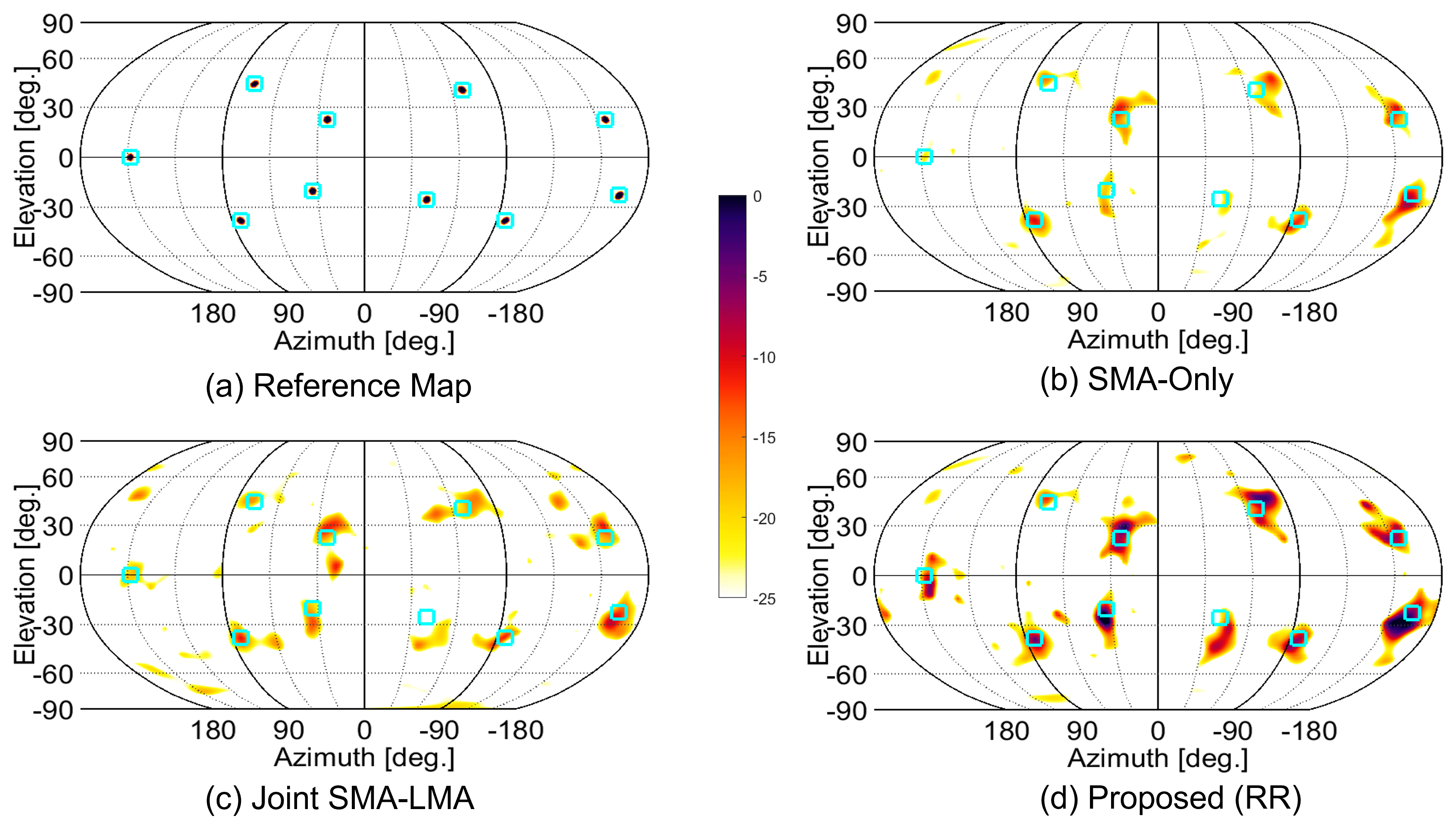}}
  \caption{Comparison of acoustic energy maps under reverberant conditions with 10 sources at a radial distance of $3.5\,\text{m}$. Color intensity reflects the accumulated energy distribution. Cyan squares indicate the true source directions.}
  \label{fig:SR example}
\end{figure}

\begin{figure*}[t]
  \centering
  \subfigure[]{\includegraphics[width=0.3\textwidth]{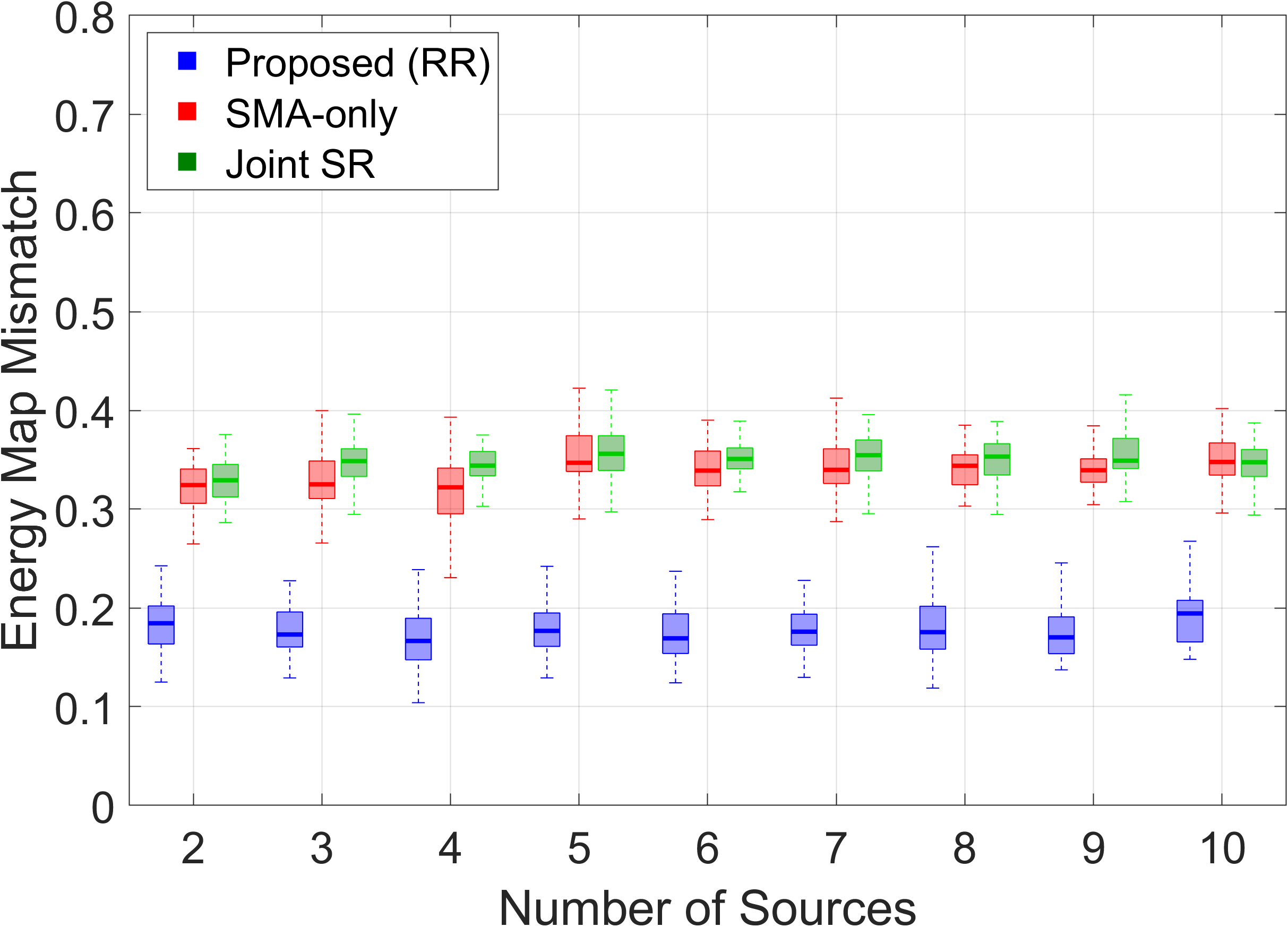}}
  \subfigure[]{\includegraphics[width=0.3\textwidth]{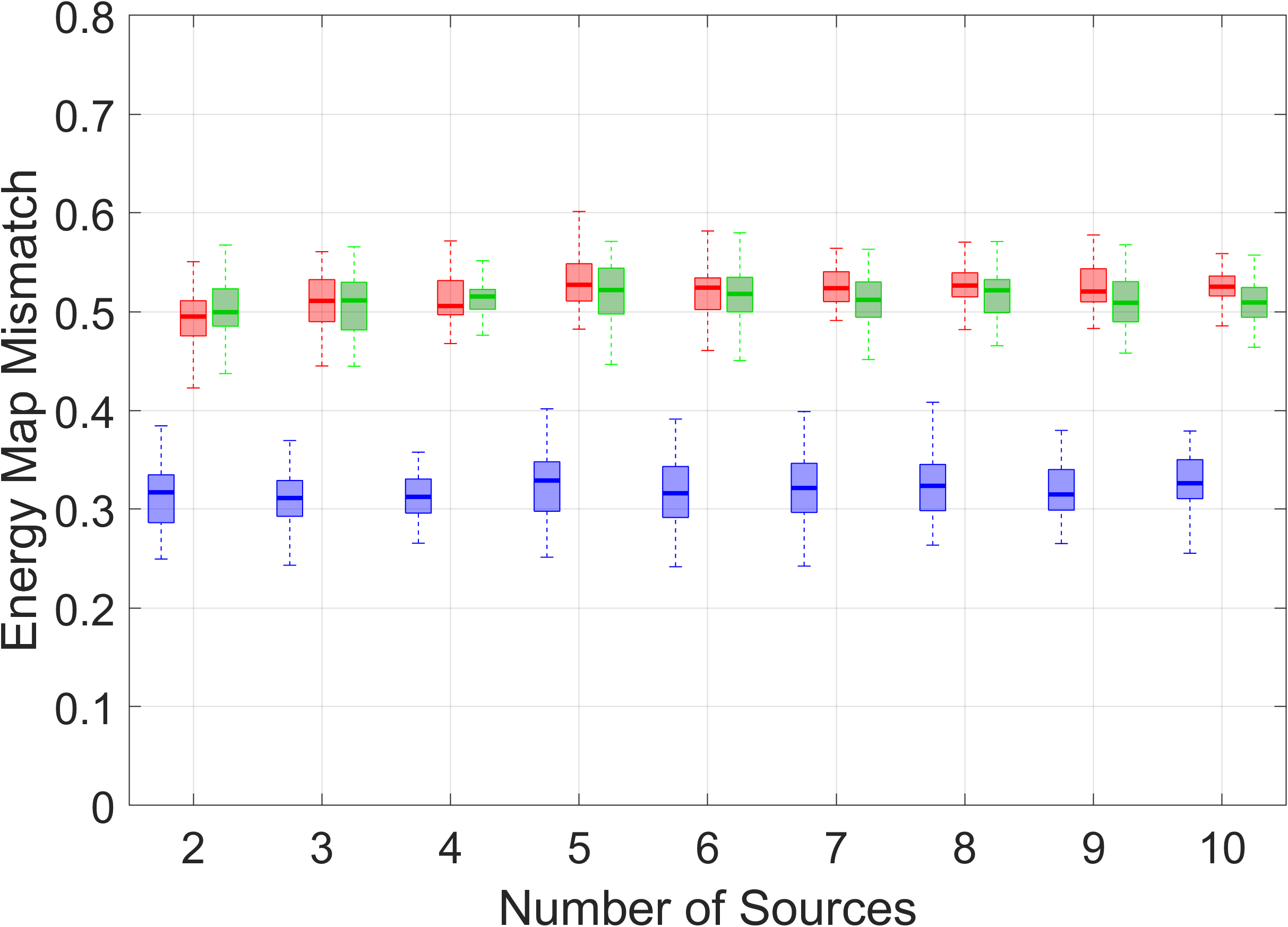}}
  \subfigure[]{\includegraphics[width=0.3\textwidth]{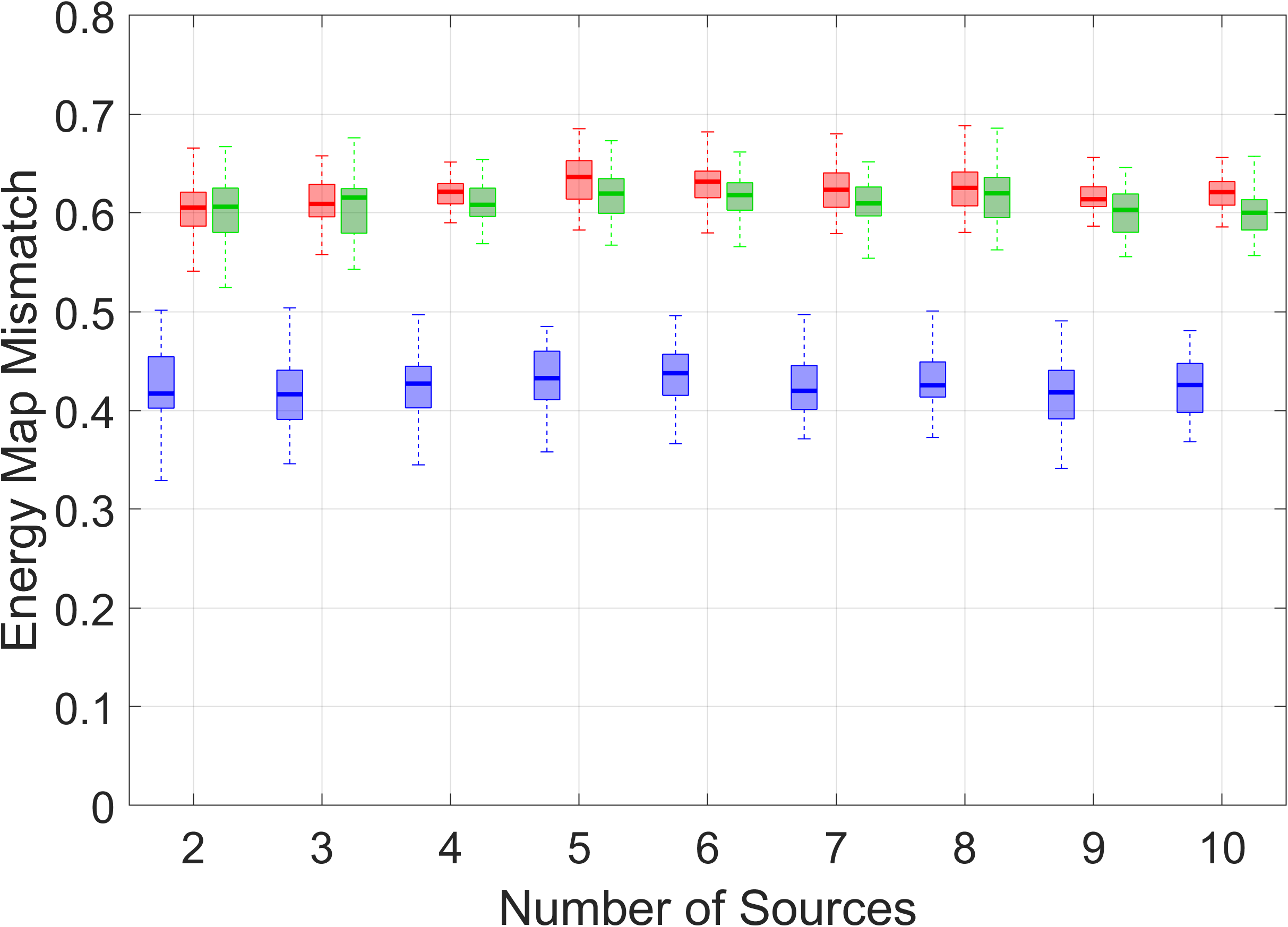}}
  \caption{Energy map mismatch across different source distances (a) 1.5m, (b) 2.5m, (c) 3.5m. Legend indicates the processing methods.}
  \label{fig:energy_mismatch room1}
\end{figure*}

\begin{figure*}[t]
  \centering
  \subfigure[]{\includegraphics[width=0.3\textwidth]{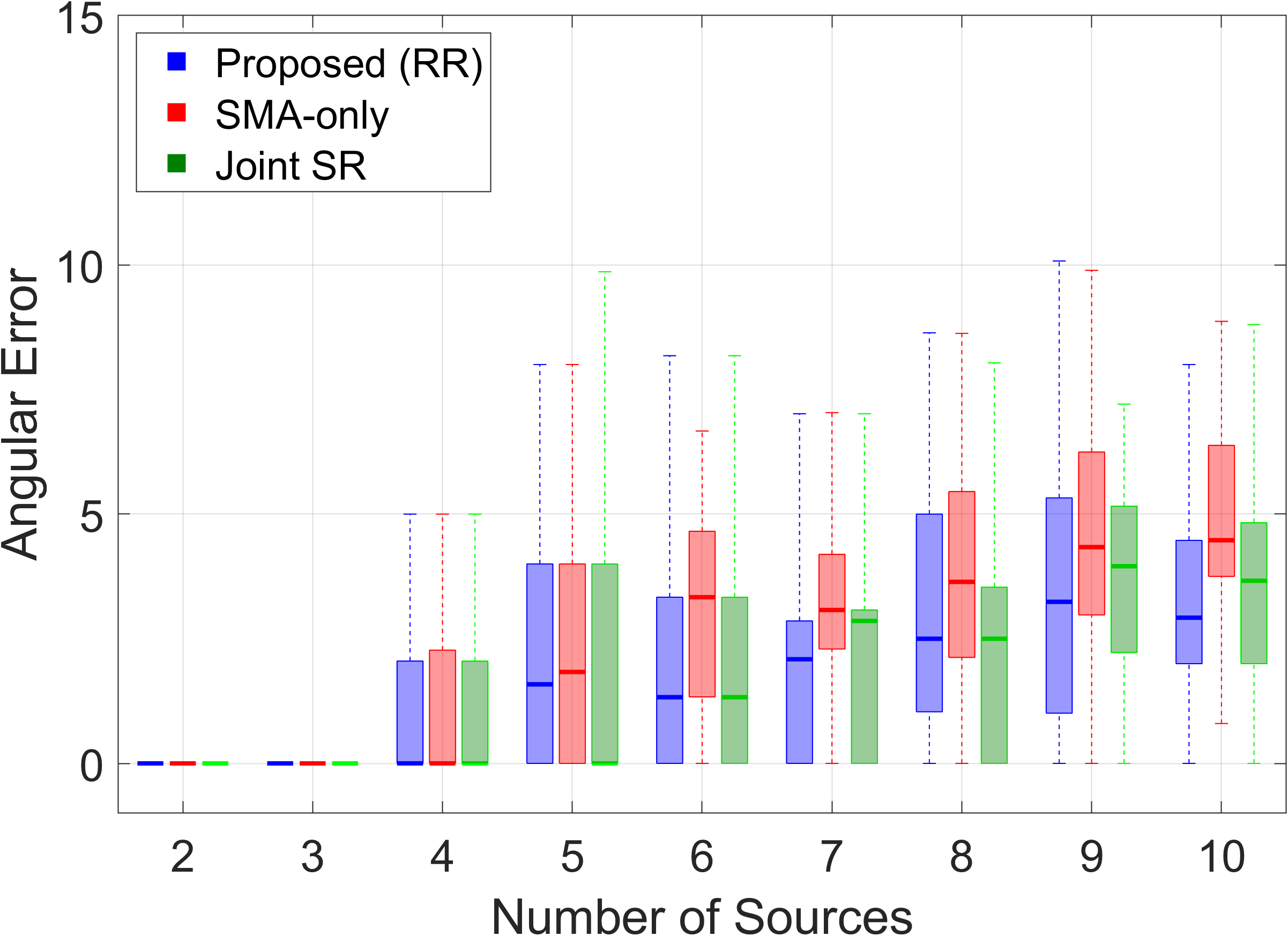}}
  \subfigure[]{\includegraphics[width=0.3\textwidth]{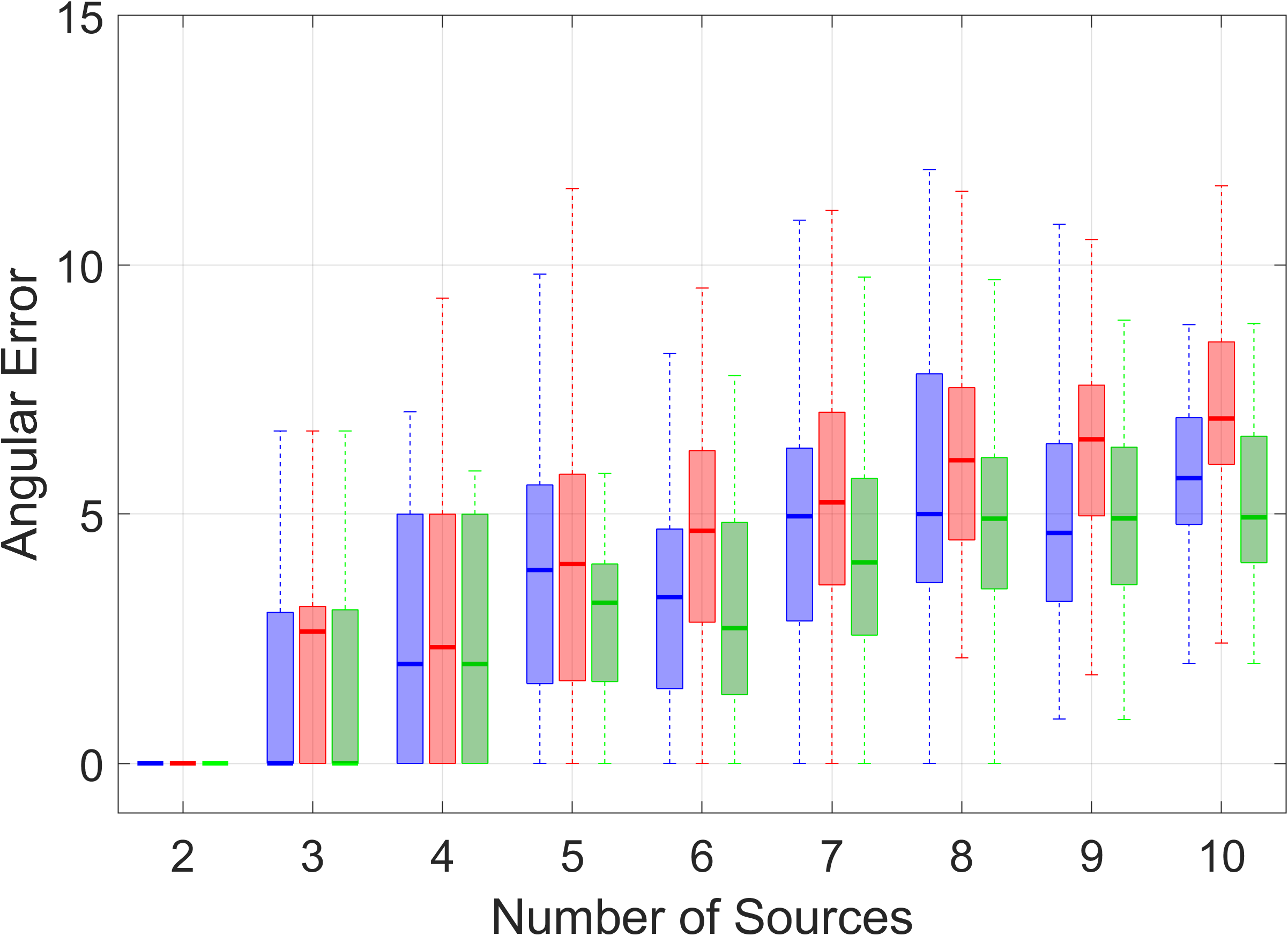}}
  \subfigure[]{\includegraphics[width=0.3\textwidth]{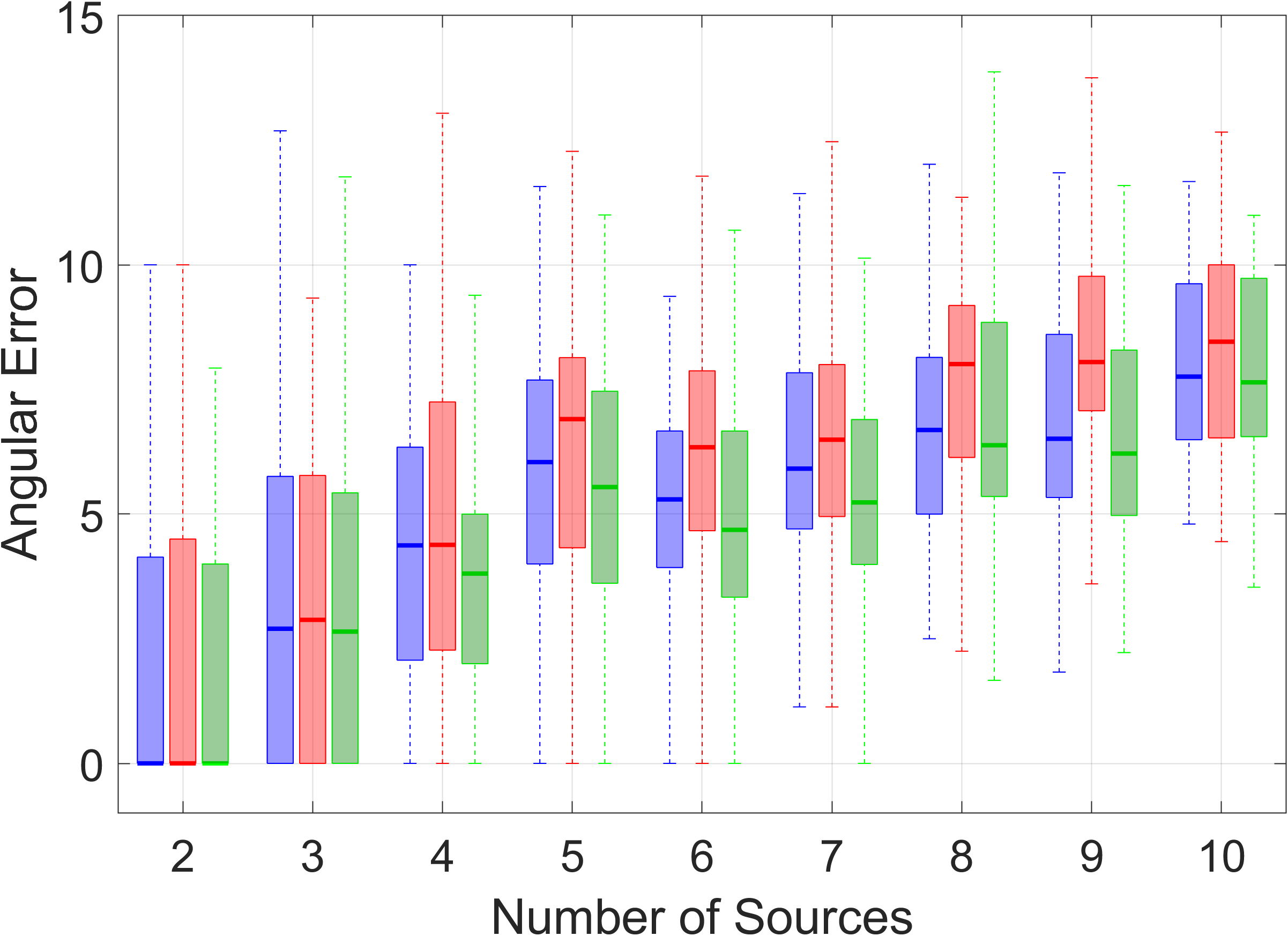}}
  \caption{Angular error across different source distances (a) 1.5m, (b) 2.5m, (c) 3.5m. Legend indicates the processing methods.}
  \label{fig:angular_error room1}
\end{figure*}

\subsection{Evaluation Metrics}
\label{subsec:metrics}
\textbf{Energy Map Mismatch:} Following \cite{jin2017sound}, the deviation between energy maps 1 and 2 is defined as:
\begin{equation}
    E = \frac{K_{11}+K_{22}-2K_{12}}{K_{11}+K_{22}},
    \label{eq2.1}
\end{equation}
where $K_{ij}$ is given by:
\begin{equation}
    K_{ij} = \sum^{Q}_{q=1}\sum^{P}_{p=1}\sqrt{\rho^{(i)}_q\rho^{(j)}_p}\,k\Big(\mathbf{\Omega}_q^{(i)},\mathbf{\Omega}_p^{(j)}\Big),
\end{equation}
where $\mathbf{\Omega}_q^{(i)}$ and $\rho_q^{(i)}$ denote the $q$-th direction in map $i$ and the corresponding power value, respectively. The function $k(\cdot,\,\cdot)$ is a spatial kernel function defined as:
\begin{equation}
    k\left(\mathbf{\Omega}_q^{(i)},\mathbf{\Omega}_p^{(j)}\right) =  \max\left(1 - \frac{\angle\left(\mathbf{\Omega}_q^{(i)},\mathbf{\Omega}_p^{(j)}\right)}{\pi/12},\ 0\right)
\end{equation}
where $\angle\left(\mathbf{\Omega}_q^{(i)},\mathbf{\Omega}_p^{(j)}\right)$ denotes the angular distance between the directions $\mathbf{\Omega}_q^{(i)}$ and $\mathbf{\Omega}_p^{(j)}$. The kernel function $k\Big(\mathbf{\Omega}_q^{(i)},\mathbf{\Omega}_p^{(j)}\Big)$ decreases linearly from 1 to 0 as the angular distance increases from 0 to $\pi/12$, and becomes zero for angular distances greater than $\pi/12$.

\textbf{Angular Error:} The angular error is defined as the angular distance between the true direction $\mathbf{\Omega}_n$ and the estimated direction $\hat{\mathbf{\Omega}}_n$:
\begin{equation}
\text{Angular Error} = \angle\big(\mathbf{\Omega}_n, \hat{\mathbf{\Omega}}_n\big).
\end{equation}
The estimated direction is selected within a $20^\circ$ neighborhood of the true direction, among those above $-20$ dB energy, and within $80\%$ of the local peak.

\subsection{Results and Discussion}
Fig.\,\ref{fig:SR example} shows an example of the acoustic energy map from three methods: SMA-only sparse recovery, one-step joint SMA-LMA sparse recovery, and the proposed two-stage SMA-LMA sparse recovery with residue refinement. The proposed method exhibits noticeably darker and more concentrated energy regions at the true source directions, indicating superior energy preservation. In contrast, the other two methods show weaker energy patterns, reflecting less accurate energy reconstruction.

Fig.\,\ref{fig:energy_mismatch room1} presents the results for energy map mismatch across the 3 methods for 3 distances ($1.5\,\text{m}$, $2.5\,\text{m}$, and $3.5\,\text{m}$): the proposed SMA+LMA hybrid with residue refinement (blue), the SMA-only method (red), and the baseline one-step SMA+LMA method (green). A lower mismatch indicates better fidelity in reconstructing the spatial energy distribution of the sound field. Across all configurations, the proposed method consistently outperforms the SMA-only and joint SR baselines, achieving the lowest median mismatch despite exhibiting slightly higher variance. The performance gap between the proposed method and the two baselines increases as the number of sources and the source distance grow, since both factors amplify reverberation effects and challenge spatial resolution. As the source distance increases, all methods show performance degradation due to increased reverberation and reduced spatial resolution, but the proposed method maintains a clear advantage in median mismatch. Notably, the straightforward one-step joint processing offers no significant benefit over the SMA-only method and can even degrade performance under reverberation conditions.

Fig.\,\ref{fig:angular_error room1} shows the results for localization errors. Both the proposed method and the joint SR baseline achieve comparable localization accuracy, consistently outperforming the SMA-only approach. Similar to the trend observed in the energy map mismatch analysis, localization performance degrades as the source distance increases. Nevertheless, the integration of LMA measurements provides additional spatial information, enabling both the proposed and joint SR methods to maintain better accuracy than the SMA-only configuration across all tested scenarios.

These results affirm that direct joint SR with mixed-array geometry does not effectively exploit the complementary properties of SMA and LMAs, particularly in reverberant conditions. Although the joint SR method achieves competitive localization accuracy, it degrades significantly in spatial energy reconstruction. In contrast, the proposed two-stage design first estimates the global 3D spatial structure using SMA data, which is robust to reverberation and well-suited for resolving elevation components. The LMAs provide high resolution in the horizontal plane but suffer from inherent elevation ambiguity. Using the LMAs only to refine the unresolved horizontal details in the residue, where any remaining elevation content is likely weak, preserves the accuracy of the 3D estimate obtained from the SMA. This design preserves the integrity of the spatial structure while allowing complementary spatial information to be incorporated.

A natural alternative for spatial energy estimation is to apply beamforming in the second stage, steering energy along the DOAs estimated from the SMA. However, this method has key limitations: it cannot reconstruct the full spatial energy map and often spreads energy into incorrect directions due to false-positive DOAs, particularly in reverberant environments. In contrast, the proposed SR-based refinement estimates the complete spatial distribution by accounting for overlapping components and interactions among dictionary atoms, offering greater robustness and effectiveness in recovering subtle energy patterns in complex acoustic scenes.

\section{Conclusion}
\label{sec:conclusion}
This work presented a two stage SR framework that integrates a central SMA with four surrounding LMAs to enhance sound field reconstruction in reverberant environments. Simulation results showed that the proposed method outperforms both the SMA-only and joint one-step SR baselines, achieving lower energy map mismatch and improved localization across varying source counts under reverberation conditions. Although slight increases in result variability were observed, the gains in spatial fidelity were substantial.

Future work will explore extending the framework to sound field interpolation tasks. Additional directions include developing a modal solution for the combined array geometry and incorporating robust relative transfer function (RTF) estimation to further enhance sparse recovery performance.










\fancyhead{}  
\printbibliography[title={References}]

\end{document}